\documentclass[prd,reprint,nofootinbib,notitlepage,aps,tightenlines,preprintnumbers,amsmath,amssymb,amsfonts,showpacs,superscriptaddress]{revtex4-1}

\usepackage{hyperref}
\usepackage{graphicx}

\usepackage{natbib,bm}
\usepackage{color}
\usepackage{multirow}
\usepackage{bbold}
\usepackage{verbatim}

\newcommand{\nc}{\newcommand}
\nc{\grad}{\nabla}  
\nc{\tr}{\mathop{\rm Tr}}
\nc{\half}{{1\over 2}}
\nc{\third}{{1\over 3}}
\nc{\be}{\begin{equation}}
\nc{\ee}{\end{equation}}
\nc{\bea}{\begin{eqnarray}}
\nc{\eea}{\end{eqnarray}}

\nc{\dint}[2]{\int\limits_{#1}^{#2}}
\nc{\D}{\displaystyle}
\nc{\PDT}[1]{\frac{\partial #1}{\partial t}}
\nc{\tw}{\tilde{w}}
\nc{\tg}{\tilde{g}}
\nc{\newcaption}[1]{\centerline{\parbox{5.6in}{\caption{#1}}}}
\def\href#1#2{#2} 

\def\beq{\begin{eqnarray}}   \def\eeq{\end{eqnarray}}
\def\s{\sigma_{\rm DW}}

\def\lsim{\mathrel{\rlap{\lower3pt\hbox{\hskip0pt$\sim$}}
    \raise1pt\hbox{$<$}}}         
\def\gsim{\mathrel{\rlap{\lower4pt\hbox{\hskip1pt$\sim$}}
    \raise1pt\hbox{$>$}}}         

\def\Z{{\mathbb{Z}}}
\def\C{{\mathbb{C}}}
\def\R{{\mathbb{R}}}

\def\Id{\hbox{1\kern-.23em{\rm l}}}

\nc{\al}{\alpha}
\nc{\ga}{\gamma}
\nc{\de}{\delta}
\nc{\ep}{\epsilon}
\nc{\ze}{\zeta}
\nc{\et}{\eta}
\renewcommand{\th}{\theta}
\nc{\Th}{\Theta}
\nc{\ka}{\kappa}
\nc{\la}{\lambda}
\nc{\rh}{\rho}
\nc{\si}{\sigma}
\nc{\ta}{\tau}
\nc{\up}{\upsilon}
\nc{\ph}{\phi}
\nc{\ch}{\chi}
\nc{\ps}{\psi}
\nc{\om}{\omega}
\nc{\Ga}{\Gamma}
\nc{\De}{\Delta}
\nc{\La}{\Lambda}
\nc{\Si}{\Sigma}
\nc{\Up}{\Upsilon}
\nc{\Ph}{\Phi}
\nc{\Ps}{\Psi}
\nc{\Om}{\Omega}
\nc{\ptl}{\partial}
\nc{\del}{\nabla}
\nc{\ov}{\overline}
\nc{\gsl}{\!\not}
\nc{\bi}[1]{\bibitem{#1}}
\nc{\fr}[2]{\frac{#1}{#2}}
\nc{\dsl}{\partial\!\!\!\!\!\!\not\,\,}
\nc{\gm}{\mbox{$\gamma_{\mu}$}}
\nc{\gn}{\mbox{$\gamma_{\nu}$}}
\nc{\Le}{\mbox{$\fr{1+\gamma_5}{2}$}}
\nc{\Ri}{\mbox{$\fr{1-\gamma_5}{2}$}}
\nc{\GD}{\mbox{$\tilde{G}$}}
\nc{\gf}{\mbox{$\gamma_{5}$}}
\nc{\Ima}{\mbox{Im}}
\nc{\Rea}{\mbox{Re}}

\nc{\av}{\langle \ph\rangle}
\nc{\ntwo}{${\cal N}\!\!=\!2\;$}
\nc{\none}{${\cal N}\!\!=\!1\;$}
\nc{\nfour}{${\cal N}\!\!=\!4\;$}

\def \bi{\bibitem}
\nc{\rf}[1]{(\ref{#1})}
\def \del{\partial}

\def\be{\begin{equation}}
\def\ee{\end{equation}}
\def\bes{\begin{equation*}}
\def\ees{\end{equation*}}
\def\ba{\begin{align}}
\def\ea{\end{align}}
\def\bea{\begin{eqnarray}}
\def\eea{\end{eqnarray}}

\def\f{\frac}

\def\v[#1]{\boldsymbol{#1}}
\def\w[#1]{\widehat{#1}}
\def\vs[#1,#2]{\boldsymbol{{#1}_{#2}}}
\def\mes[#1]{d^{3}{#1}}
\def\del{\partial}
\def\<{\langle}
\def\>{\rangle}
\def\vec[#1]{\boldsymbol{#1}}
\def\vecs[#1,#2]{\boldsymbol{{#1}_{#2}}}

\newcommand{\B}[1]{{\bar{1}}}
\newcommand{\BD}[1]{{\dot \bar{1}}}

\def\a{\alpha}
\def\b{\beta}

\def\D{\Delta}
\def\e{\epsilon}

\def\G{\Gamma}

\def\l{\lambda}
\def\L{\Lambda_2}
\def\m{\mu}
\def\n{\nu}

\def\O{\Omega}
\def\p{\phi}

\def\s{\sigma}
\def\S{\Sigma}

\def\th{\theta}

\begin{document}

\title{Domain wall moduli in softly-broken SQCD at \boldmath{$\bar\theta=\pi$}}

\author{Adam Ritz}
\email{aritz@uvic.ca}
\author{Ashish Shukla}
\email{ashish@uvic.ca}
\affiliation{Department of Physics and Astronomy, University of Victoria, 
Victoria, BC V8P 5C2, Canada}


\begin{abstract}
\noindent We analyze the moduli space dynamics of domain walls in $SU(N)$ QCD at $\bar\theta=\pi$, by softly breaking ${\cal N}\! =\!1$ SQCD with sfermion mixing. In the supersymmetric limit, BPS domain walls between neighbouring vacua are known to possess non-translational flavour moduli that form a $\C {\rm P}^{N-1}$ sigma model. For the simplest case with gauge group $SU(2)$ and $N_f=2$, we show that this sigma model also exhibits a Hopf term descending from the bulk Wess-Zumino term with a quantized coefficient. On soft-breaking of supersymmetry via sfermion mixing that preserves the flavour symmetry, these walls and their moduli-space dynamics survives when $\bar\theta=\pi$ so that there are two degenerate vacua.
\end{abstract}

\maketitle



\section{Introduction}
\label{Intro}
QCD with gauge group $SU(N)$ and $N_f$ massive quark flavours explicitly violates $T$-symmetry unless the physical theta-angle $\bar\theta=0$ or $\pi$. Current evidence suggests that real-world QCD lies very close to $\bar\theta=0$, but it is nonetheless interesting to explore the physics near $\bar\theta=\pi$. In particular, at large $N$ the $\bar\theta$-dependence of the vacuum energy conjecturally takes the form $E_0(\bar\theta) \propto N^2 {\rm min}_k f((\bar\th+2\pi k)/N)$ for some multi-branched function $f$ \cite{Witten:1980sp,dvv,Ohta,Witten:1998uka}. This implies that when $\bar\theta=\pi$, at large $N$ the $T$-symmetry is spontaneously broken and there are two degenerate vacua with $k=0,N-1$, suggesting the existence of domain walls interpolating between them. 

This vacuum structure has recently been clarified for finite $N$ by the identification of a mixed discrete 't Hooft anomaly for $T$-symmetry and a 1-form center symmetry \cite{gkks} (See also \cite{gks,DVRVY,TK}). The presence of this anomaly, along with the assumption of a gapped spectrum, provides a more general argument for the spontaneous breaking of $T$-symmetry and the presence of degenerate vacua at $\bar\theta=\pi$. (See \cite{Creutz,Smilga} for earlier work.) It follows that domain walls should exist interpolating between these two degenerate vacua, and arguments were provided that these walls should possess nontrivial flavour moduli. Namely, for gauge group $SU(N)$, it has been argued that the worldvolume theory should be a 2+1D $\C {\rm P}^{N_f-1}$ sigma model with a Wess-Zumino term \cite{gks}.

This picture can be made quite concrete in \none SQCD with $N_f=N$, albeit in a regime in which the $\th$-parameter is no longer physical. This theory 
has a low energy description on the Higgs branch, in terms of meson and baryon chiral superfield moduli $M$, $B$ and $\tilde{B}$, 
where it reduces to a massive perturbation of a K\"ahler sigma model on the
manifold determined by the quantum constraint \cite{Seiberg}
\be
 {\rm det}M - B\tilde{B} = \La_N^{2N}\, . 
 \label{mod}
\ee
The massive theory possesses $N$ degenerate quantum vacua (see Fig.~1), consistent with the Witten index, with a range of interpolating BPS domain walls \cite{ds,kss,sv,ksy,dCM,dk,av,rsv1,rsv2}. The additional supersymmetry ensures that this system can be studied analytically in the weakly-coupled regime where one quark flavour is parametrically heavier than the others. We review this construction in the next section, but an important conclusion is that the BPS $k$-walls, connecting vacua between which the vacuum value of the superpotential changes phase by $2\pi k/N$, come in multiplets associated with the vacua of a worldvolume sigma model. 
In \cite{rsv1,rsv2} it was argued that $k$-walls exhibit a nontrivial classical reduced moduli space $\widetilde{\cal M}_k$ due to localized Goldstone modes associated with the
flavour symmetries which are broken by the wall solution. The corresponding coset is a complex Grassmannian \cite{rsv1},
\be
 \widetilde{\cal M}_k = G(k,N) \equiv 
 \frac{{\rm U}(N)}{{\rm U}(k) \times {\rm U}(N-k)} \ . 
 \label{Gr}
\ee
Note that for $k=1$, $G(1,N)=\C {\rm P}^{N-1}$.  One can then formally deduce that the multiplicity of $k$-walls, $\nu_k$, is given by
the worldvolume Witten index for this Grassmannian sigma model, which depends only
on the topology of the space, and is given by the Euler characteristic, $\nu_k ~=~ \ch(G(k,N)) ~=~ \frac{N!}{k!(N-k)!}$. 

\begin{figure}[t]
\centerline{\includegraphics[viewport=150 540 460 730, clip=true, scale=0.4,width=0.35\textwidth]{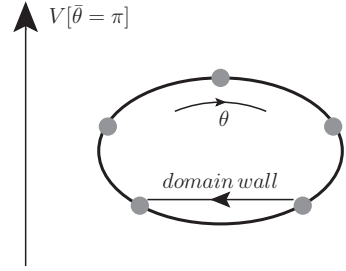}}
\vspace*{0cm}
 \caption{\footnotesize A schematic representation of the perturbation to the $N$ degenerate vacua of SQCD (for $N=5$), which creates a potential tilting the plane, and for $\bar\theta=\pi$ results in two degenerate vacua which allow for interpolating domain walls.}
\end{figure}

In this paper, we first consider the extension of this picture due to the presence of a bulk Wess-Zumino term in the chiral theory describing meson and baryon fields \cite{Manohar:1998iy, Dubovsky:1999hv}. For the simplest case of gauge group SU(2), i.e. $N=2$, we show that the presence of this term leads to a parity-odd Hopf term in the moduli space Lagrangian for the $\C {\rm P}^1$ sigma model. Furthermore, we perturb this picture by softly breaking \none SUSY by adding an $F$-term spurion for the squark mass \cite{ess}, which renders the $\th$-parameter physical once more, and generically lifts the degeneracy between the vacua. The perturbation is tractable in the regime $F_m \ll m \ll \La$ and leads to a potential,
\be
 \De V = - {\rm Re} {\rm Tr} (F_m M) = -N_f |F_m| |M| \cos\left(\frac{\bar\th + 2\pi n}{N}\right).
\ee
This manifests two degenerate vacua at $n=0,N-1$ in the special case that $\bar\th=\pi$, suggesting the possible presence of a 1-wall (using the terminology of $k$-walls above) connecting them (see also \cite{fz} and the recent analysis in \cite{Draper}). We use this tractable perturbation to examine the worldvolume moduli of this domain wall, that is inherited from the unperturbed BPS 1-walls in \none SQCD. Most importantly, this perturbation does not break any of the residual flavour symmetries if $F_m$ is diagonal, and thus the domain wall $\C P^{N-1}$ flavour moduli and the associated Hopf term should persist in this regime.

In the next two sections, we review the analysis of domain wall moduli arising from broken flavour symmetries in SQCD, and then show that the presence of a bulk Wess-Zumino term leads in the $SU(2)$ case to a Hopf term in the moduli space Lagrangian for walls separating neighbouring vacua. In section \ref{sec:three} we turn on the soft-breaking $F$-term perturbation and analyze the residual moduli space dynamics for walls connecting the degenerate vacua for $\bar\th=\pi$. We end with some concluding remarks in section \ref{conclusions}.

\section{Domain Wall Moduli in \boldmath{\none} SQCD}
\label{dwmoduli}
In this section we briefly review the arguments 
which determine the topology of the reduced $k$-wall moduli space in SU($N$) 
SQCD with $N_f=N$ massive flavours \cite{rsv1}.

\subsection{Vacuum structure and flavour symmetry}

\none SQCD with $N_f=N$ flavours is obtained by adding $N$ chiral superfields,
$Q_{f}$ and $\widetilde {Q}^{\bar g}$ ($f,{\bar g}=1,\ldots,N$), transforming respectively in the
fundamental and anti-fundamental representations of the gauge group, to the fields
of \none SYM with gauge group SU($N$). This matter content will ensure that the
gauge symmetry is completely broken in any vacuum in which the matter fields
have a nonzero vacuum expectation value. Provided the mass gap is sufficiently large,
the gauge fields may then be integrated out, obtaining a low energy effective
description in terms of the meson and baryon moduli $M_f^{\bar g} = Q_f \widetilde{Q}^{\bar g}$,
$B=\ep^{f_1 f_2\cdots f_N} Q_{f_1}Q_{f_2}\cdots Q_{f_N}$, and $\tilde{B} = \ep_{\bar{g}_1\bar{g}_2\cdots \bar{g}_N} \widetilde{Q}^{\bar{g}_1}\widetilde{Q}^{\bar{g}_2}\cdots \widetilde{Q}^{\bar{g}_N}$.

The superpotential describing the resulting low energy dynamics is given by \cite{Seiberg}
\be
 {\cal W} = {\rm Tr}(\hat{m}M) + \la \left({\rm det}M-B\tilde{B}-\La_N^{2N}\right), 
  \label{sp}
\ee
in terms of the meson matrix $M$, the baryon fields $B$, $\tilde{B}$, the dynamical scale $\La_N$, and
a Lagrange multiplier $\la$. The Lagrange multiplier is to be understood as a 
heavy classical field, for consistency with the nonrenormalization theorem, which
enforces the quantum constraint ${\rm det}M-B\widetilde{B} = \La_N^{2N}$ \cite{Seiberg} shown in (\ref{mod}) (see also \cite{is,sv1}).

With degenerate masses, the vacua preserve a maximal $SU(N)$ flavour symmetry. However, to ensure that the vacua and generic interpolating domain wall trajectories will lie at weak coupling, the quark mass matrix $\hat m$ needs to be hierarchical. This ensures that the gauge modes, which have been integrated out, are indeed heavy relative to the strong coupling scale $\La_N$. The choice which retains the maximal $SU(N-1)$
global symmetry is given by
\be
 \hat{m} = {\rm diag}\{m, m ,\ldots, m, m_N \}\ , \;\;\;\;\;\;\; 
   \La_N \gg m_N \gg m\ . \label{hmm}
\ee
This leads to $N$ degenerate vacua, given by diagonal meson vacuum expectation values
(VEVs) with components (no summation over $i$),
\be
 \langle M_{i}^{i} \rangle_n = \left(\frac{m_N}{m}\right)^{1/N}\!\!\La_N^2 \om_N^n\,, 
 \quad \om_N^n={\rm e}^{2\pi i n/N}\,, \label{vac}
 \ee
where $i=1,\ldots,N-1$ and $n=0,\ldots,N-1$, while $\langle B\rangle = \langle \tilde{B} \rangle =0$.
The vacua are weakly coupled if the hierarchy is sufficiently large: i.e. if $m_N/m \gg e^N$. Restricting attention to energy scales below $m_N$, the 
effective dynamical scale is $\La_{N-1}^{2N+1}=m_N\La_N^{2N}$.

Despite this hierarchical setup, an important observation of \cite{rsv1,rsv2} that will be relevant here is that we can restore the full $SU(N)$ flavour symmetry in the superpotential by rescaling fields and, since the breaking of flavour symmetry is only visible in the Ka\"hler potential, the spectrum of BPS domain walls will be independent of the rescaling. As discussed in more detail below, this nonrenormalization theorem will allow us to effectively focus on the regime with maximal $SU(N)$ flavour symmetry.

\subsection{BPS $k$-walls and flavour moduli}

The $N$ vacua discussed above are conveniently represented in the complex plane of the superpotential ${\cal W}$, where they lie on a circle about the origin,
\be
 \left. {\cal W} \right|_{\rm vacua} \longrightarrow {\cal W}_n = N({\rm det} \, \hat{m})^{1/N} \La_N^2 {\rm e}^{2\pi n/N}.
\ee
The presence of degenerate vacua allows for domain wall solutions interpolating between them. A class of these domain walls are 1/2-BPS, preserving half of the supersymmetry of the vacuum. BPS walls inherit certain nonrenormalization properties, with a tension $T_k$ given by a central charge in the supersymmetry algebra, $T_k = |{\cal Z}_{kn}|$, where  ${\cal Z}_{kn} = 2({\cal W}_{n+k} - {\cal W}_{n})$. The tension is determined by the vacuum values of the superpotential only, and moreover the walls themselves are described by straight lines between the vacua in the ${\cal W}$ plane \cite{fmvw,cv}. To analyze the generic features of the wall trajectories, we can set $B=\tilde{B}=0$ since they vanish in the vacua, although they will be relevant when we consider the moduli space dynamics later on. It is then convenient to define dimensionless fields
$X=\hat{m}M(\mu\La_N^2)^{-1}$, with $\mu\equiv ({\rm det}\,\hat{m})^{1/N}$,
in terms of which the superpotential exhibits the maximal SU($N$) flavour symmetry,
\be
 {\cal W} = \mu\La_N^2 \left[ {\rm Tr}\,X + \la({\rm det}X - 1)\right],
   \label{spX}
\ee
while the hierarchical structure of the mass matrix is now visible
only in the rescaled K\"ahler potential. The superpotential depends only
on the eigenvalues  $\{\et_i\}$ of $X$,
\be 
 {\cal W} = \mu\La_N^2 \left[ \sum_{i=1}^N \et_i +
 \la\left(\prod_{i=1}^N \et_i - 1\right)\right],
\ee
which exhibits the vacua at $\langle \et_i \rangle_k = \om_N^k$. With $k$-wall boundary conditions, so that the wall interpolates between vacua separated by a phase difference of $e^{2\pi i k/N}$, the trajectory of each eigenvalue is determined by its winding number $w(\et)$ in the complex ${\cal W}$ plane, namely either $w_1=k/N$ or $w_2=k/N-1$ (see also \cite{hiv}). The Bogomol'nyi equations satisfied by 1/2-BPS walls then ensure that $N-k$ of the eigenvalues carry winding
number $w_1$ and $k$ carry winding number $w_2$. 

This structure implies that $k$-walls fall into multiplets,  corresponding to the various winding trajectories of the eigenvalues. One way to understand this structure, developed in \cite{rsv1,rsv2}, is to consider the moduli space of the wall configurations associated with symmetries that are spontaneously broken by the wall configuration.  On the general grounds that a $k$-wall spontaneously 
breaks translational invariance, the moduli space can be decomposed as follows,
\be
 {\cal M} = \R \times \widetilde{\cal M},
\label{decomp}
\ee
where the factor $\R$ reflects the center of mass position $z_0$, while
$\widetilde{\cal M}$ denotes the reduced moduli space. Furthermore, it follows from the 
constraints on the eigenvalues that
the maximal flavour symmetry the $k$-wall can preserve is
\be
 {\rm SU}(k) \times {\rm SU}(N-k) \times {\rm U}(1)\ ,
\ee
which is a subgroup of the full flavour symmetry SU($N$).
Consequently, accounting for discrete symmetries, there must be
localized Goldstone modes on the wall parametrizing the Grassmannian
coset $\widetilde{\cal M}_k = G(k,N) \equiv {\rm U}(N)/({\rm U}(k) \times {\rm U}(N-k))$ \cite{rsv1,rsv2}, shown in (\ref{Gr}).
In particular, for minimal 1-walls between neighbouring vacua, we have
\be
 \widetilde{\cal M}_1 = \C {\rm P}^{N-1},
\ee
associated with the flavour symmetries broken by the wall.

This moduli space will be the focus of the next section. For completeness, we recall that the nonrenormalization properties of BPS states mean that the $k$-wall multiplicity is given by an index, specifically the CFIV index \cite{cfiv,cv} that counts shortened 1/2-BPS multiplets. It is defined as the following trace, suitably regularized, 
over the Hilbert space with boundary conditions appropriate to a 
$k$-wall \cite{cfiv,cv}, $\nu_k \equiv {\rm Tr}\, F (-1)^F$,
where $F$ is the fermion number operator. This index reduces to the worldvolume Witten index ${\rm Tr}\,(-1)^F$ \cite{Witten1} of the Grassmannian sigma model, which is given by the Euler characteristic. Thus, in the presence of a suitable infrared regulator, the multiplicity of $k$-walls is given by \cite{rsv1,rsv2},
 \be
 \nu_k ~=~ \ch(G(k,N)) ~=~ \frac{N!}{k!(N-k)!}.
\ee
This result is fully consistent with the number of permutations of the eigenvalues $\et_i$, given the constraints on the winding number discussed above \cite{rsv1}. Moreover, it shows that the result depends only on the topology of the reduced moduli space of BPS walls.
In particular, it is only the induced metric on this space which is sensitive to the precise
specification of quark masses; the topology is invariant. Since the index is independent of smooth diffeomorphisms of 
the K\"ahler potential \cite{cfiv}, we can restore its symmetry by such
a diffeomorphism if so desired.  The worldvolume theory can be analyzed more explicitly for $N=N_f=2$, as we discuss in the next section.

\section{Dynamics of 1-wall moduli for \boldmath{$N=N_f=2$}}
\label{dwdynamics}
This structure can be verified explicitly for the gauge group SU(2), where the reduced meson constraint ${\rm det}M = \La_2^{4}$ defines the 6-dimensional manifold $T^*(S^3)$, and the two vacua lie at the poles of the base $S^3$. Direct analysis of the BPS equations then shows \cite{rsv1,rsv2} that the wall profile lies entirely within an $S^1$ fibre of the $S^3$, allowing fluctuations of the remaining meson fields parametrizing the $S^2 = \C {\rm P}^1$ to form the worldvolume flavour moduli, consistent with the symmetry argument above.

We will utilize the K\"ahler deformation argument above in order to work within the fully flavour symmetric regime.\footnote{The impact of the hierarchical structure of the mass matrix is discussed in \cite{rsv1,rsv2}, and amounts to the generation of a potential on the moduli space.} However, we will need to retain the full space of moduli including the baryonic directions,
\be
\label{mbcons}
\text{det} M - B \tilde{B} = \L^4\, ,
\ee
for reasons that will become clear when we consider the presence of a bulk Wess-Zumino term. For completeness, we recall that the meson matrix $M^a_{\bar{b}} = Q^a \widetilde{Q}_{\bar{b}}$, where $Q^a$ and $\widetilde{Q}_{\bar{b}}$ are the quark and anti-quark fields respectively. The baryon and anti-baryon fields are $B = \e_{ab} Q^a Q^b$ and $\widetilde{B} = \e^{\bar{a}\bar{b}} \widetilde{Q}_{\bar{a}} \widetilde{Q}_{\bar{b}}$. Note that since we are working with $N_f = N_c =2$, we have $a, \bar{a} = 1,2$ etc.

We can rewrite the constraint eq.\eqref{mbcons} in terms of six complex scalar fields $X_1, \cdots X_6$ as
\be
\label{mbcons2}
\sum_{i=1}^6 X_i^{\,2} = 1,
\ee
with the identifications $M_{11} = \L^2 (X_1 + i X_2)$, $M_{22} = \L^2 (X_1 - i X_2)$, $M_{12} = \L^2 (i X_3 + X_4)$, $M_{21} = \L^2 (i X_3 - X_4)$, $B = \L^2 (i X_5 + X_6)$, and $\tilde{B} = \L^2 ( iX_5 - X_6)$.

BPS walls also satisfy first order Bogomol'nyi equations \cite{fmvw,cv},
\be
 g_{\bar{i} j} \ptl_z X^j = e^{-i\gamma} \ptl_{\bar{i}} \bar{\cal W},
\ee
where the K\"ahler metric is given by $g_{\bar{i}j} = \ptl_{\bar{i}} \ptl_j {\cal K}$, while $\ptl_{\bar{i}} = \ptl/\ptl \bar{X}^{\bar i}$, $\ptl_j = \ptl/\ptl X^j$, and $\ga$ is the phase of the central charge ${\cal Z}_{kn}$.

The K\"ahler metric on the space spanned by the mesonic and baryonic moduli is unknown, except at weak coupling where it is inherited from the canonical kinetic terms for the squark fields. As discussed above, the vacua and wall trajectories lie at weak coupling with $M_{f}^{\bar g} \gg \La_2^2$ only if the (s)quark mass matrix is hierarchical. To preserve the full flavour symmetry in the superpotential, we have rescaled the fields, reducing the symmetry of the K\"ahler metric. However, as reviewed above, the domain wall moduli are determined topologically, and are largely independent of the choice of the K\"ahler metric, provided that it is non-singular. In the equal mass regime with $m_1=m_2=\mu$, the two vacua now lie at the poles of the $S^3$ which forms the real section of the surface $\sum_{i=1}^4 X_i^2=1$ (with $X_5=X_6=0$). Moreover, as discussed in \cite{rsv1,rsv2}, the full wall trajectory lies within this $S^3$ subspace, and in the limit that the (s)quark mass perturbation is switched off, the enhanced $SU(2)\times SU(2)$ flavour symmetry implies that the leading dependence on a non-canonical K\"ahler metric can be reabsorbed within the Bogomol'nyi equations into a rescaling of the transverse coordinate $z$. The re-introduction of a hierarchical (s)quark mass matrix was shown to lead to a potential in the domain wall moduli space which will not concern us here.

With these arguments in mind, we now proceed to study the domain wall moduli space while making use of a canonical K\"ahler metric for the meson and baryon moduli $\{X_{i=1\ldots6}\}$, subject to the constraint (\ref{mbcons2}). In practice, the wall trajectories only involve the real section of this constraint which is an $S^5$, for which 
a simple hyperspherical coordinate system will be convenient to use,
\be
\begin{split}
 X_1 &= \cos\ps \\
 X_2 &= \sin\ps \cos\xi \\
 X_3 &= \sin\ps \sin\xi \cos\ph \\
 X_4 &= \sin\ps \sin\xi \sin\ph \cos\al \\
 X_5 &= \sin\ps \sin\xi \sin\ph \sin\al \cos\th \\
 X_6 &= \sin\ps \sin\xi \sin\ph \sin\al \sin\th,
\end{split}
\label{S5}
\ee
where all angles other than $\th$ run from $0$ to $\pi$, while $0<\th\leq 2\pi$.
Notice that on setting $\al=0$, the parametrization (\ref{S5}) reduces to a coordinatization of an $S^3/\Z_2$ submanifold,
\be
\begin{split}
 X_1 &= \cos\ps \\
 X_2 &= \sin\ps \cos\xi \\
 X_3 &= \sin\ps \sin\xi \cos\ph \\
 X_4 &= \sin\ps \sin\xi \sin\ph \\
 X_5 &= X_6 =0,
\end{split}
\label{S3}
\ee
with the metric
\be
 ds^2_{S^3/\Z_2} = d\ps^2 + \sin^2\ps \left( d\xi^2 + \sin^2\xi \,d\ph^2\right). \label{2met}
\ee
In practice, after setting the baryon fields to zero, the real section of $\sum_{i=1}^4 X_i^2=1$ is a full $S^3$. It is an artefact of the hyperspherical coordinate system that only $S^3/\Z_2$ is covered by setting $\al=0$. This feature of the coordinates (\ref{S5}) will not play a significant role in the present subsection, as it only impacts global properties of the moduli space, but it will be important when we consider the Wess-Zumino term below. Note that the other half of the full meson field space, a second $S^3/\Z_2$ submanifold, is obtained by setting $\al=\pi$. 

The $S^3/\Z_2$ submanifold at $\al=0$ covered by the coordinate system above will be sufficient to (locally) construct the geometry of the wall moduli space. The superpotential reduces to
\be
 {\cal W} = {\rm Tr} (\hat{m}M) =  2\mu \La_2^2 X_1 \longrightarrow 2 \mu \La_2^2 \cos\ps,
\ee
with the vacua at the poles $\ps=0,\pi$ of the 3-sphere. Within this coordinate system, the Bogomol'nyi equations take the form
\be
 \ptl_z \ps = - 2\mu \sin\ps\,, \qquad \ptl_z \xi = \ptl_z \ph =0\ ,
\ee
which can be solved by the sine-Gordon soliton \cite{rsv1,rsv2},
\be
 \label{sgsol}
 \begin{split}
 \ps_{\rm sol}(z) &= 2\arctan\left(e^{-2\mu(z-z_0)}\right), \\
   \xi_{\rm sol}&=\xi_0,\;\;\;\;\;\;  \ph_{\rm sol}=\ph_0\ . 
   \end{split}
\ee

We can now compute the bosonic moduli space explicitly, by integrating the meson and baryon kinetic terms $\frac{1}{2}\La_2^2 \ptl_\mu X_i \ptl^\mu X_ i$ over the transverse direction $z$ to the wall. Setting $\al=0$ to remain in the $S^3/\Z_2$ subspace, the result is a worldvolume action for the wall moduli \cite{rsv1,rsv2} 
\be
  S_{\cal M} = \frac{1}{2} \int d^3 x  \left[ T_1\, \partial_{\mu} z_0\, \partial^{\mu}z_0+ h_{ij} \ptl_\mu x^i \ptl^{\mu}x^j\right], \label{Lmod}
\ee
where $T_1=4\mu\La_2^2$ is the 1-wall tension, and $h_{ij}$ is the
metric on the flavour moduli space, given by \cite{rsv1,rsv2},
\begin{align}
 ds^2_{{\cal M}} &= T_1\, d z_0^2 +h_{ij}dx^idx^j \\
  &= T_1\, d z_0^2 + R_{\widetilde{\cal M}} \left( d\xi_0^2 + \sin^2\xi_0 d\ph_0^2\right), \nonumber
\end{align}
with $R_{\widetilde{\cal M}} = \frac{\La_2^2}{\mu}$ the scale of the flavour moduli space. It follows that ${\cal M}^{\rm N=2} = {\R} \times {\C}{\rm P}^1$, consistent with the general symmetry arguments above, although our coordinates only cover half the flavour moduli space $\C {\rm P}^1/\Z_2$.

\subsection{Wess-Zumino term in $SU(2)$ SQCD}
As in QCD, SQCD with $N_f=N$ has a low energy symmetry realization that requires the presence of a Wess-Zumino term in the chiral Lagrangian describing meson and baryon moduli. The real section of the moduli space constraint ${\rm det}M - B\tilde{B} = \L^4$ topologically describes a  5-sphere $S^5$, and so following Witten \cite{Witten:1983tw}
a Wess-Zumino term can be written down to ensure the correct symmetry under parity, that has a quantized coefficient which follows by considering this term as a map from a 5D-disk with 4D spacetime as the boundary to $S^5$. 

The relevant Wess-Zumino term was first written down for gauge group $SU(2)$ with $N_f=2$ by Manohar \cite{Manohar:1998iy}, with an extension to gauge group $SU(N)$ proposed in \cite{Dubovsky:1999hv},\footnote{The representation here follows \cite{Dubovsky:1999hv}. It matches with the original expression given by Manohar \cite{Manohar:1998iy} up to an overall sign within the coordinate parametrization \eqref{S5}.}
\begin{align}
\label{wz1}
\G_{SU(2)} &= -\,\f{1}{12\pi^2 \L^8} \,\text{Im} \int_{\S_5} d\O\; \text{det} M\, \e^{\m\n\l\rho\s} \del_{\m}B \, \del_\n\tilde{B} \nonumber\\ 
 &\qquad\times \text{Tr}\left( M^{-1} \del_\l M \,M^{-1} \del_\rho M \, M^{-1} \del_\s M \right).
\end{align}
Note that the coefficient is fixed by quantization, while $\m, \n, \cdots$ run over $(t,x,y,z,w)$, where $(t,x,y,z)$ are the usual 4D spacetime coordinates and $w$ denotes an additional coordinate on the 5D disk used to define the Wess-Zumino term.

Transforming from $M$, $B$ and $\tilde{B}$ to the fields $X_1, \cdots X_6$, as defined above, we will be interested in the real section of the constraint eq.\eqref{mbcons2}, obtained by demanding that $X_1,\cdots X_6$ are real scalars. We will utilize the hyperspherical parametrization (\ref{S5}), with spacetime coordinate dependence adapted to the wall configuration and moduli, namely $\ps=\ps(z)$ is a function of the transverse coordinate only while $(\xi,\ph,\th)$ will be taken as functions of the wall worldvolume coordinates $(t,x,y)$. Recalling that the wall flavour moduli arise from the $S^3$ submanifold, we will define the remaining angle $\al=\al(w)$, a function only of the fifth dimensional coordinate running from $w=0$ to the 4D spacetime boundary at $w=1$. Requiring that $\al(w=1)=0$ in 4D spacetime ensures that only the meson moduli $X_1,\cdots X_4$ are excited along the wall, consistent with the discussion in the previous subsection. However, with $\al\neq0$ within the bulk of the 5D disk, the non-vanishing baryon moduli effectively support the Wess-Zumino term. 

Substituting the parametrization (\ref{S5}) into the meson/baryon kinetic terms with the above coordinate dependence (including $\al(w=1)=0$) and integrating over $z$ leads again to the moduli space Lagrangian (\ref{Lmod}), while the Wess-Zumino term (\ref{wz1} reduces to
\be
\begin{split}
\Ga^{(\rm wall)}_{SU(2)} =  - \, \f{2}{\pi^2} &\int_{\S_5} d\O \sin^4\psi \, \del_z \psi \sin \a \, \del_w \a\\
&\times \sin^3 \xi \sin^2 \phi \, \e^{\m\n\rho} \, \del_\m \th \, \del_\n \phi \, \del_\rho \xi.
\end{split}
\ee
Note that the integral over the additional disk coordinate $w$ takes the form $\int d\al \sin\al$, with the integrand being symmetric about the midpoint where $\al=\pi/2$. Since the coordinates only cover half the mesonic subspace at $\al=0$ (or at $\al=\pi$), for consistency we only integrate over half the disk, $\int_0^{\pi/2} d\al \sin\al = 1$. On carrying out the remaining integral over $z$, the Wess-Zumino term becomes
\be
\label{hopf2}
  \Ga^{(\rm wall)}_{SU(2)} \rightarrow S_{\rm H} = -\, \f{3}{4 \pi} \int d^3 x \sqrt{G} \, \e^{\m\n\rho} \, \del_\mu \th_0 \, \del_\n \p_0 \, \del_\rho \xi_0,
\ee
where 
\be
\label{m3rg}
 \sqrt{G} = \sin^3 \xi_0 \sin^2 \ph_0
\ee
defines the restriction of the $S^5$ volume measure to the compact 3D submanifold ${\cal M}_3$ spanned by $\{\xi,\ph,\th\}=\{\xi_0,\ph_0,\th_0\}$, which are moduli for the wall solution. ${\cal M}_3$ is a fibration of $S^1$ parametrized by $\th$ over the base $\C {\rm P}^1$ associated with the wall flavour moduli. This is consistent with the conventional structure of the $\C {\rm P}^1$ linear sigma model, in which the reduction from a round $S^3$ to $\C {\rm P}^1$ is achieved by gauging the $U(1)$ fibre.

The structure of (\ref{hopf2}) is recognizable as the Hopf term for the $\C {\rm P}^1$ sigma model \cite{wz,prs,bds} (as reviewed in the Appendix). Rotating to Euclidean space, and compactifying spacetime to an $S^3$, we can identify the (Euclidean) spacetime coordinates with the target space coordinates $\{\xi,\ph,\th\}$ as the simplest embedding, and integrate the Hopf measure over the $S^3$. This leads to a quantization condition for the Hopf map $S^3 \rightarrow \C {\rm P}^1$ in the form $-\frac{3i}{4\pi}\int_{S^3} \sqrt{G} d\xi d\ph d\th = - i\pi$. As reviewed in the Appendix, the quantization condition on the right hand side has the generic form $ i \Th_H \times \Z$, where $\Th_H$ (which has period $2\pi$) is the coefficient of the normalized Hopf term. We conclude that in this model
\be
\label{hopfth}
 \Theta_H = \pi,
\ee
reflecting the underlying breaking of parity associated with the bulk Wess-Zumino term. Interestingly, while $\Th_H$ is naively a periodic real parameter, it has recently been argued that there is a further quantization condition, such that only $\Th_H =0$ and $\pi$ are consistent at the quantum level \cite{fks}. We will comment on this in the final section \ref{conclusions}.

\section{Soft breaking to QCD}
\label{sec:three}
One of our primary goals was to understand how the moduli space structure of domain walls elaborated above responds under soft breaking of supersymmetry. A convenient perturbation to turn on is the $F$-term component of the squark mass matrix $\hat{m}$, treating the mass as a background superfield \cite{ess},
\be
 \De V = - {\rm Re} {\rm Tr} (F_m M).
\ee
This correction to the scalar potential is a perturbation provided that $|F_m| \ll m \ll \La_N$, and importantly will not modify the flavour symmetry if we assume that $F_m$ is aligned in flavour space with the (s)quark mass matrix $\hat{m}$. Making this assumption, so that as complex matrices $F_m = \ep\hat{m}$, for gauge group $SU(N)$ with $N_f=N$ we have
\be
 \De V = - {\rm Re}\left[ \ep \mu \La_N^2 {\rm Tr} (X) \right],
\ee
where $X$ is again the rescaled meson matrix. As shown in (\ref{vac}), the meson vevs prior to the perturbation take the form $\langle X \rangle_n \propto e^{2\pi n/N}$. The introduction of the spurion $F_m$ explicitly breaks the $U(1)_R$ symmetry, rendering the $SU(N)$ $\th$-parameter physical. The physical value $\bar\th$ is a linear combination of the bare parameter $\th_0$, which multiplies the tr$(G\tilde G)$ term in the action, and the phases of $\hat{m}$ and $F_m$ \cite{ess}. We can use the anomalous $U(1)_A$ symmetry and the broken $U(1)_R$ symmetry to rotate all these into the phase of $F_m$. As a perturbation to the potential $V$, it follows that in vacuum, 
\be
 \De V = -N |\ep\mu \La_N^2| \cos\left(\frac{\bar\th + 2\pi n}{N}\right), 
\ee
which generically lifts the vacuum degeneracy, leaving a unique vacuum state for one value of $n$. However, in the special case that $\bar\th=\pi$ as noted in the Introduction (and shown in Fig.~1), the system manifests degenerate (neighbouring) vacua. Although supersymmetry is broken, it follows on general topological grounds that the SQCD 1-walls between these vacua will survive the perturbation. Moreover, since this $F$-term perturbation does not break any further flavour symmetries, the wall will continue to inherit $\C {\rm P}^{N-1}$ flavour moduli. Thus, in the regime  $|F_m| \ll m \ll \La_N$, the full structure of the bosonic moduli space dynamics for 1-walls, as discussed in the previous sections, should translate directly to this softly broken regime. In contrast, the domain walls between other vacua will be metastable once the relevant vacuum states are split. 

Although softly broken SQCD in the confining phase is expected to be continuously connected to QCD, it is important to keep in mind that the degrees of freedom are distinct. In the QCD regime, $|F_m|\gg \La_N$, the flavour moduli are presumably realized differently. Moreover, we anticipate that the fermionic moduli discussed in \cite{rsv2} will be lifted by this perturbation as they are no longer protected by the presence of spontaneously broken supercharges. However, their mass should be parametrically low in this limit, so that the world volume theory would be distinct from the one for domain walls in pure QCD at $\th=\pi$.

\section{Concluding Remarks}
\label{conclusions}
In this short paper, we have presented a concrete test of the recently discussed discrete symmetry realization in QCD at $\bar\theta=\pi$ \cite{gkks,gks}. Spontaneous breaking of $T$-symmetry allows for domain walls interpolating between two degenerate vacua, and moreover the worldvolume dynamics of these walls should break parity via the presence of a Wess-Zumino term. We have approached this question in the present paper by considering \none SQCD, which possesses a spectrum of BPS domain wall solutions between degenerate vacua, with worldvolume moduli associated with broken flavour symmetries. We have argued that for SQCD with gauge group $SU(2)$ and $N_f=2$, the worldvolume theory for walls connecting neighbouring vacua, a $\C {\rm P}^1$ sigma model in 2+1D, also exhibits a Hopf term which descends from the bulk Wess-Zumino term. When \none supersymmetry is softly broken by an $F$-term deformation that preserves the same flavour symmetry, the worldvolume theory for 1-walls is preserved at $\bar\th=\pi$, providing a concrete, and positive, test of the expected structure in the QCD limit. 

In the remainder of this section, we comment on some implications of these results, and potential extensions.

\begin{itemize}
\item {\it Quantization conditions}:- The quantization of the 4D Wess-Zumino term in SQCD translates to a specific quantization condition for the coefficient $\Th_H$ of the Hopf term on the wall worldvolume. As shown in section \ref{dwdynamics}, we obtain $\Th_H=\pi$ in the $SU(2)$ case with $N_f=2$. Recently, such a quantization condition has been considered in \cite{fks}, where it was argued that under certain conditions the only consistent values of $\Th_H$ are $0$ and $\pi$. The present example with an explicit microscopic realization apparently satisfies this more restrictive criterion. 

\item {\it Extension to $SU(N)$}:- We have analyzed the simplest scenario with gauge group $SU(2)$, but the result should generalize to higher rank gauge groups as well. For SQCD with gauge group $SU(N)$ the spectrum of $k$-walls expands, but it is clear that on soft-breaking of supersymmetry, only the 1-walls between neighbouring vacua can survive the perturbation when there are two degenerate vacua at $\bar\th=\pi$. The wall flavour moduli space then becomes $\C {\rm P}^{N-1}$, for which a Hopf term also exists. The linear 
sigma model Lagrangian, generalizing the discussion in the Appendix, takes the form
\begin{align}
\label{cpN}
\qquad\quad  {\cal L}_{CP^{N-1}} &= \frac{1}{f} \left(\del_\m Z^*_\a \, \del^\m Z_\a - (Z^*_\a \del_\m Z_\a)(Z_\b \del^\mu Z^*_\b)\right) \nonumber\\
 &+ \l \left(\sum_{\a = 1}^N |Z_\a|^2 - 1\right) -\f{\Theta}{4\pi^2} \e^{\m\n\rho} A_\mu \ptl_\nu A_\rh,
 \end{align}
where $Z_\al$, $\al=1,\ldots,N$, defines a complex $N$-vector of fields, while the $U(1)$ gauge field takes the form $A_\mu = \frac{1}{2}\left( Z^*_\a \del_\m Z_\a{-}Z_\a \del_\mu Z^*_\a \right)$. It would be interesting to verify whether this term arises from the reduction of the proposed $SU(N)$ Wess-Zumino term \cite{ds} to the wall worldvolume. This would be consistent with the proposed structure in QCD for $N_f=N$ \cite{gks}. 

\item {\it Wall worldvolume supersymmetry}:- As a final comment, we note that within SQCD itself the presence of the higher-derivative Hopf term on the wall worldvolume may also clarify the worldvolume realization of supersymmetry. As discussed in \cite{rsv2}, the K\"ahler structure of the flavour moduli space leads to an enhanced \ntwo supersymmetry. Although this does not extend to the full moduli space including the translational coordinate, this extended supersymmetry is surprising as it is not protected by the BPS structure of these states. The appearance of the Hopf term suggests that this symmetry enhancement is accidental at the two-derivative level, and not preserved at higher-derivative order. The Hopf term can be realized within the supersymmetric $\C {\rm P}^1$ model \cite{hs}, so it would be interesting to verify this explicitly. 
\end{itemize}
 
\acknowledgments{We would like to thank M. Shifman and A. Vainshtein for many helpful discussions over the years about BPS states in supersymmetric gauge theories. This work was supported in part by NSERC, Canada.}

\appendix 

\section{The Hopf term in the $ \C {\rm P}^1$ sigma model}
In this appendix, we briefly review some features of the $\C {\rm P}^1$ model with a Hopf term \cite{wz,prs,bds,Hong:2001vk}). The model is defined in 2+1 spacetime dimensions, with a Lagrangian density given by
\be
\label{lang1}
{\cal L} = {\cal L}_{CP^1} + {\cal L}_{\rm H},
\ee
where 
\begin{align}
\label{cp1}
{\cal L}_{CP^1} &= \frac{1}{f} \left(\del_\m Z^*_\a \, \del^\m Z_\a - (Z^*_\a \del_\m Z_\a)(Z_\b \del^\mu Z^*_\b)\right) \nonumber\\
 &\qquad + \l \left(|Z_1|^2 + |Z_2|^2 - 1\right),
\end{align}
where $\a = 1, 2$, $\l$ is a Lagrange multiplier whose equation of motion ensures the constraint $|Z_1|^2 + |Z_2|^2 = 1$, and $f$ is a dimensionful coupling that sets the scale of $\C {\rm P}^1$. The Hopf term is conveniently viewed as a Chern-Simons Lagrangian for the non-dynamical $U(1)$ vector field $A_\mu = \frac{1}{2}\left( Z^*_\a \del_\m Z_\a{-}Z_\a \del_\mu Z^*_\a \right)$,
\begin{align}
\label{hopf}
{\cal L}_{\rm H}&{=} -\f{\Theta}{4\pi^2} \e^{\m\n\rho} A_\mu \ptl_\nu A_\rh \nonumber\\
 &{=}-\f{\Theta}{8\pi^2} \e^{\m\n\rho} \left( Z^*_\a \del_\m Z_\a{-}Z_\a \del_\mu Z^*_\a \right) \del_\n Z^*_\b \, \del_\rho Z_\b .
\end{align}

A parametrization for the complex fields $Z_\a$ that automatically meets the constraint is
\be
\begin{split}
\label{para1}
&Z_1 = e^{i\,(\th + \p)/2} \, \text{sin}\left(\f{\xi}{2}\right),\, Z_2 = e^{i\,(\th - \p)/2} \, \text{cos}\left(\f{\xi}{2}\right).
\end{split}
\ee
Here $\th, \p, \xi$ are functions of $(t,x,y)$ such that $\th \in [0,4\pi), \p \in [0, 2\pi)$ and $\xi \in [0, \pi]$. 

Substituting this parametrization into \eqref{cp1} we find a 2+1D sigma model on $\C {\rm P}^1$,
\be
\begin{split}
\label{cp1p}
{\cal L}_{CP^1} = 
\frac{1}{4f} \left[ \ptl^\mu \xi \ptl_\mu \xi + \sin^2\xi\, \ptl^\mu \p \ptl_\mu \p \right],
\end{split}
\ee
where $\mu =(t,x,y)$, and we note that there is no dependence on $\th$ in \eqref{cp1p}. Similarly, substituting the parametrization \eqref{para1} into the Hopf term \eqref{hopf} gives
\be
\label{hopf3}
{\cal L}_{\rm H} = -\, \f{\Theta}{2 \pi^2} \sqrt{G} \, \e^{\m\n\rho} \, \del_\mu \th \, \del_\n \p \, \del_\rho \xi ,
\ee
where $\sqrt{G}$ is the volume measure for the 3-manifold with coordinates $\{\xi,\ph,\th\}$,
\be
 \sqrt{G} = \frac{1}{8} \sin\xi.
\ee

Rotating to Euclidean space, and compactifying spacetime to an $S^3$, we can identify the (Euclidean) spacetime coordinates with the target space coordinates $\{\xi,\ph,\th\}$ as the simplest embedding, and integrate the Hopf measure over the $S^3$. This leads to a quantization condition for the Hopf map $S^3 \rightarrow \C {\rm P}^1$ in the form $-\frac{i\Th}{2\pi^2}\int_{S^3} \sqrt{G} d\xi d\ph d\th = - i\Th$, indicating that $0 \leq \Th < 2\pi$ is a periodic angle. Recent work \cite{fks} suggests that in fact only $\Th=0$ and $\Th=\pi$ are fully consistent at the quantum level.

\bibliography{refs}

\end{document}